\documentclass[sigconf,nonacm]{acmart}

\usepackage{hyperref}

\usepackage{algorithmic}
\usepackage{graphicx}
\usepackage{textcomp}
\usepackage{tcolorbox}
\usepackage{enumitem}
\usepackage{booktabs}
\usepackage{pifont}
\usepackage{microtype}

\usepackage{multirow}

\usepackage{makecell}

\usepackage{graphicx}

\usepackage[table]{xcolor}
\usepackage{subcaption}
\definecolor{lightyellow}{RGB}{255, 253, 240}
\definecolor{plaincol}{RGB}{210,170,160}
\definecolor{levelcol}{RGB}{170,185,210}
\definecolor{skillcol}{RGB}{225,205,180}

\begin{document}

\definecolor{paleyellow}{rgb}{1, 1, 0.95}
\definecolor{lower}{rgb}{0.88235,0.7451,0.41569}
\definecolor{lower}{rgb}{0.94, 0.85, 0.6}
\definecolor{higher}{rgb}{0.5, 0.8, 0.75}
\definecolor{rowColor}{rgb}{0.937, 0.937, 0.937}
\definecolor{mauve}{rgb}{0.25,0,0.52}
\definecolor{darkerteal}{RGB}{1, 77, 78} 
\newcommand{\FIXME}[1]{\textcolor{red}{Revision: \uline{#1}}}
\newcommand{\fixme}[1]{#1}
\newcommand{\MLE}[1]{\textcolor{blue}{MLE: #1}}
\newcommand{\mle}[1]{\textcolor{blue}{MLE: #1}}
\newcommand{\erfan}[1]{\textcolor{red}{Erfan: #1}}
\newcommand{\Erfan}[1]{\textcolor{red}{Erfan: #1}}
\newcommand{\veronica}[1]{\textcolor{green}{Veronica: #1}}
\newcommand{\Veronica}[1]{\textcolor{green}{Veronica: #1}}
\newcommand{\peggy}[1]{\textcolor{orange}{Peggy: #1}}
\newcommand{\Peggy}[1]{\textcolor{orange}{Peggy: #1}}
\newcommand{\Chris}[1]{\textcolor{brown}{Chris: #1}}
\newcommand{\chris}[1]{\textcolor{brown}{Chris: #1}}

\newcommand{\subheading}[1]{\vspace{2pt}\noindent\textbf{#1}}
\newcommand{\qualquote}[2]{%
  \textit{\textcolor{mauve}{``#1}''} \hspace{0em}%
  \textnormal{(#2)}%
}

\DeclareRobustCommand{\code}[1]{%
  \textit{\textcolor{darkerteal}{#1}}%
}
\definecolor{grey}{rgb}{0.5,0.5,0.5}
\newcommand{\GroupOne}[0]{\section{Internal Problem-Solving}}
\newcommand{\GroupTwo}[0]{\section{Hitting Targets}}
\newcommand{\GroupThree}[0]{\section{Environmental Disruptions}}
\newcommand{\GroupFour}[0]{\section{Job Fit and Social Aspects}}

\definecolor{borderblue}{RGB}{0,102,204}

\newcommand{\summarytcolorbox}[2]{%
  \par\vspace{0.25cm}%
  \noindent
  \makebox[\columnwidth][l]{%
    \fcolorbox{mauve!75!black}{borderblue!5!white}{%
      \begin{minipage}{\dimexpr\columnwidth-2\fboxsep-2\fboxrule\relax}
        \textbf{\textit{#1}:} #2
      \end{minipage}%
    }%
  }%
  \par
}


\newenvironment{quoteblock}%
  {\begin{list}{}{\leftmargin=2em \rightmargin=1em \topsep=5pt \parsep=0pt \itemsep=8pt} \item[]}%
  {\end{list}}

\definecolor{upcellbg}{HTML}{EAF4FC}     
\definecolor{downcellbg}{HTML}{FCECEC}   

\newcommand{\upcell}[1]{%
  \cellcolor{upcellbg}\makebox[0pt][r]{$^{\uparrow}$\,}\textbf{#1}%
}
\newcommand{\downcell}[1]{%
  \cellcolor{downcellbg}\makebox[0pt][r]{$^{\downarrow}$\,}\textbf{#1}%
}

\title{Generative AI May Reinforce Social Biases in Software Engineering Education}


\author{Erfan Entezami}
\affiliation{%
  \institution{University of Massachusetts Amherst}
  \city{Amherst, MA}
  \country{USA}}
\email{eentezami@cs.umass.edu}

\author{Andrew Lan}
\affiliation{%
  \institution{University of Massachusetts Amherst}
  \city{Amherst, MA}
  \country{USA}}
\email{andrewlan@cs.umass.edu}

\author{Madeline Endres}
\affiliation{%
  \institution{University of Massachusetts Amherst}
  \city{Amherst, MA}
  \country{USA}}
\email{mendres@umass.edu}








\begin{abstract}

Generative artificial intelligence (GenAI) is increasingly being deployed across a wide range of real-world applications. Without careful evaluation, reliance on these systems can have unintended consequences, such as reinforcement of stereotypes and amplification of social biases. Such risks are particularly important in educational settings, where early decisions can shape students’ interests, opportunities, and career trajectories. In this paper, we investigate how GenAI use by software engineering instructors may inadvertently reinforce software-specific social biases. We focus on two representative tasks: team formation based on student profiles and the generation of visual content for educational materials. Our results reveal significant biases in both tasks. In team formation, factors such as gender and nationality affect role assignments (e.g., women are more likely to be assigned to front-end roles than equally qualified men). In the visual content generation task, models generally produce diverse and balanced representations when depicting groups of individuals. However, when generating images of a single person, the outputs are predominantly male and light-skinned. These findings highlight the persistence of demographic biases in generative models and underscore the need for domain-specific evaluation and mitigation strategies to support their responsible use in educational settings.
\end{abstract}



\maketitle

\section{Introduction}

With the rapid advancement of artificial intelligence (AI), particularly GenAI, a wide range of AI-powered tools have been developed to support teaching and learning~\cite{wang2026large, wang2026learnmate2, looi2025personalization}. These tools are increasingly being adopted for educational tasks, such as using large language models (LLMs) to provide student feedback~\cite{stamper2024enhancing}, automate grading~\cite{liu2026ai}, and deliver personalized tutoring~\cite{looi2025personalization}. Similarly, image generation models are being used to create educational materials, including lecture and presentation slides~\cite{ali2024picture, bian2025effects}. As GenAI becomes more integrated into educational settings, it is increasingly important to evaluate how these systems behave and respond when interacting with students, teachers, and education-related topics. Leveraging GenAI models in educational settings without a comprehensive understanding of their behavior can be particularly concerning as they can influence learning experiences, shape educational and career perceptions, and have long-term societal impacts. These risks highlight the need to carefully evaluate their behavior and potential biases before adopting them for educational content generation or decision-making.

Since GenAI models are trained on massive internet-scale datasets, they can inherit and reproduce biases present in their training data, leading to systematic disparities in their outputs. Prior research has shown that both text and image generation models can reinforce stereotypes related to gender, race, and other demographic attributes~\cite{AYOUB2024186, kotek2023gender, naik2023social, rozado2026gender}. While these biases are well documented, they can also emerge in domain-specific settings, where hidden patterns in training data are often more subtle and difficult to detect.

One such domain is software engineering, which has historically exhibited significant gender imbalance. Prior work has documented instances of bias in GenAI models in domain of software engineering. For example, Parziale et al.~\cite{parziale2026once} have shown that LLMs make biased recruitment and task assignment decisions based on candidates' country and pronouns rather than technical qualifications. Similarly, Bano et al.~\cite{bano2025does} examine both text and image generation models and show that, across both modalities, male and Caucasian candidates are favored over individuals from other demographic groups.

This imbalance is also reflected in educational settings, where men continue to constitute the majority of computer science students. According to the 2025 Taulbee Survey~\cite{cra2025taulbeeBachelors}, approximately 74\% of bachelor's students in computer science are men, compared with 24\% women in the U.S. Despite this persistent gender imbalance, its potential impact on GenAI models used for educational tasks remains largely unexplored. This raises concerns that GenAI  models may learn and perpetuate existing biases when generating educational content or interacting with students in software engineering education.

In this work, we explore whether, and to what extent, GenAI models exhibit social bias and discrimination toward students in the context of software engineering education. We focus on team formation and visual content generation as two underexplored applications of GenAI models in software engineering education. Through these tasks, we investigate whether text and image generation models reinforce demographic biases in educational settings.

For the team formation task, we evaluate three LLMs, with each model generating 11,200 team assignment decisions for synthetic student profiles. Using a controlled experimental design, we systematically vary the qualifications of the students (e.g. technical skills) and the personal attributes (e.g. gender, nationality) to examine how these factors influence the decisions of the model. For the visual content generation task, we analyze 500 images produced by two Text-to-Image models 
using prompts extracted from real software engineering course materials.

By combining qualitative and quantitative analysis methods, we show that state-of-the-art GenAI models exhibit biases toward specific genders, nationalities, and skin tones in educational tasks. This pattern aligns with prior findings on imbalances in the software engineering community~\cite{parziale2026once, bano2025does, rodriguez2021perceived}.

\section{Background and Related Work}

Training LLMs on large-scale, imbalanced datasets with limited supervision can lead to models that learn and reinforce social biases and stereotypes~\cite{hofmann2024ai, kotek2023gender}. Gender bias in LLMs has been extensively studied with models shown to often generate responses aligned with gender stereotypes~\cite{kotek2023gender, rozado2026gender, wan2023kelly, omiye2023large}. Racial and nationality biases are similarly well documented, with models exhibiting disparate behaviors across demographic groups~\cite{hofmann2024ai, omiye2023large, weissburg2025llms, kamruzzaman2025exploring}.

Bias in LLMs extends beyond race and gender to other demographic factors, including caste and religion~\cite{seth2025deep, abrar2025religious}, culture~\cite{tao2024cultural, dai2025word}, occupation~\cite{gorti2024unboxing, iso2025evaluating, zhang2025hire}, cognitive factors~\cite{malberg2025comprehensive, echterhoff2024cognitive, sumita2025cognitive} and recommended products~\cite{kamruzzaman2024global, kelly2025understanding, zhang2025invisible}. Collectively, these studies demonstrate that LLM biases are pervasive and vary across domains, making general mitigation challenging.

As LLMs become more prevalent in education, exploring their performance for educational tasks is increasingly important. Prior work has examined LLM biases in educational contexts and their impact on learners. \cite{weissburg2025llms, lee2024life, zheng2025dissecting}. Zheng~\cite{zheng2025dissecting} investigates bias in college major recommendations, focusing on demographic disparities (e.g., race, gender, and socioeconomic status). Weissburg et al.~\cite{weissburg2025llms} examine bias in the generation and selection of educational content, showing that LLM outputs vary across dimensions including race, gender, disability, income, and nationality. Overall, these findings show that AI biases can affect educational outcomes. However, various educational tasks such as team formation and visual content generation remain underexplored.

Prior work has raised similar concerns about the biases and social impact of LLMs in software engineering, proposing frameworks for evaluating bias and discrimination across the software lifecycle \cite{spiegler2025images, morales2024dsl, buscemi2025mind, morales2025imagebite}. 

Much of this research focuses on code generation. For example, prior work has investigated bias in LLM-based code completion and code generation tasks, including bias-sensitive prompts and mitigation strategies~\cite{liu2023uncovering, huang2025bias}.
Other work introduces evaluation frameworks for fairness in generated code~\cite{ling2025bias, du2025faircoder}, or examines how factors such as language influence model behavior~\cite{wang2024exploring}. Zhang et al. show that LLMs may systematically favor certain service providers when generating code that relies on external APIs~\cite{zhang2025invisible}.

More recently, several studies examine fairness in broader software engineering tasks~\cite{bano2025does, treude2023she, parziale2026once, nakano2024nigerian, mastropaolo2025triumph}. For example, Bano et al.~\cite{bano2025does} investigate bias in hiring using synthetic developer personas, while Nakano et al.~\cite{nakano2024nigerian} identify regional biases in role assignment using real-world GitHub profiles. Treude et al.~\cite{treude2023she} demonstrate gender bias in task descriptions through translation  analysis.

Despite growing research on AI bias in software engineering, prior work has largely focused on narrow tasks and individual bias dimensions, with limited attention to educational contexts.
Closely related work by Parziale et al.~\cite{parziale2026once} studies bias in software team composition and task assignment using explicit attributes such as country and gender. We extend prior work by using a structured framework to define software development team roles, focusing on educational settings by incorporating factors such as class standing, examining how implicit attributes (e.g., names and activities signaling gender or nationality) influence software-related decision-making, and exploring AI-generated visual content for educational materials in software engineering education.

\section{Methodology \& Experiment Design}

\subsection{Task 1: Team Formatting}
\label{subsec_TeamFormatting}

Instructor-led team formation is an important yet challenging task in educational settings, where emerging AI systems have the potential to support and streamline the process.~\cite{parker2019launching}. We define a task in which an LLM assigns a list of $n$ students to roles in a team-based software engineering course, simulating AI-assisted team formation.
Teams are derived from software engineering roles defined in the Software Engineering Body of Knowledge (SWEBOK) \cite{washizaki2024guide}. We define four teams: \textit{Interface Design}, \textit{Core Development}, \textit{Database}, and \textit{Quality Assurance}, and prompt the LLM to assign each student to the team where they are expected to contribute most effectively. Although the task generalizes to classes of arbitrary size, our experiments use classes of $n = 28$ students ($7$ per team). This configuration reflects a realistic class size while ensuring balanced demographic representation across student personas.

We construct a corpus of student personas, each contains \textit{personal attributes} and, \textit{ qualifications}. In our persona dataset, \textit{personal attributes} are surfaced via a name that may indirectly signal a student's gender and nationality. Ideally, such signals should not influence the LLM's team assignments. Importantly, these attributes are not stated explicitly; they can only be inferred indirectly from the name. \textit{Qualifications}, on the other hand, are designed to reflect technical skills and experiences that may be plausibly relevant to team placement.

\paragraph{Persona Personal Attributes.}

We desired a set of names that reflected a diverse set of genders and nationalities. For our \textit{personal attributes}, we selected names commonly associated with various countries and genders so that they serve as indirect signals of the student's demographic attributes. To select target countries, we referenced the Open Doors report \cite{opendoors_origin}, an annual survey conducted by the Institute of International Education that provides data on international student enrollment in the United States. In addition to nationalities representing North American backgrounds, such as the U.S. and Canada, we select six of the most common nationalities among international students in the U.S., prioritizing diversity in language and culture. Ultimately, we select seven countries: India, China, South Korea, Vietnam, Nepal, Nigeria, and the United States.

Next, for each nationality, we select 10 male-associated and 10 female-associated names from the most common names reported by Forebears~\cite{forebears_names}, a genealogy database. This process resulted in a dataset of 140 names drawn from seven country groupings with balanced gender representation.

\paragraph{Persona Qualifications.}
\label{subsec:qual}

For the \textit{qualifications} component, we consider three different scenarios representing instructors having varying levels of information about students’ technical capabilities and skills. We evaluate all three strategies to examine how the availability of such information may influence the model’s final decisions. The motivation for using multiple strategies is that instructors may have different levels of access to student information.

\paragraph{Strategy 1: Plain.}
We do not provide any information about the students’ skills. As a result, each persona only contains the \textit{personal attributes} component, which is the student's name. This level of information is readily available to instructors through the course roster.

\noindent\paragraph{Strategy 2: Level-based.}
We include each student’s year in a four-year undergraduate program (freshman through senior) as an indicator of their academic progress and experience. Course instructors might have access to this information through the university's student information system.

\paragraph{Strategy 3: Skill-based.}
Each persona includes a description of technical skills (e.g., programming languages, tools, libraries, and domain knowledge) derived from 12 relevant LinkedIn job postings for the four target teams. From these postings, we construct representative skill sets reflecting real-world software engineering requirements and combined them into six skill-based profiles. Each combination aligns a student with two potential teams, enabling us to study how LLMs make assignment decisions when multiple suitable options exist. This strategy reflects scenarios in which instructors use additional information, such as survey responses or self-reported skills and preferences, to form student teams more effectively, consistent with the team-matching engineering literature~\cite{parker2019launching}.

To assess the influence of personal attributes on model decisions, we generate multiple versions of each class by remapping names to different qualifications, ensuring that no name is paired with the same qualification more than once. We create four versions for the \textit{Skill-based} strategy, three for the \textit{Level-based} strategy, and one for the \textit{Plain} strategy, which includes only student names.

Each class version is evaluated using GPT-4.1, GPT-5.2, and DeepSeek V3.2. To account for the non-deterministic nature of LLMs, each version is evaluated over 10 runs with a randomly shuffled ordering of student personas. In every run, models are instructed to assign students to teams that maximize their effectiveness for a programming project based on the provided information. A summary of the structure of the experiments and the total number of runs for the team formation task is provided in Table~\ref{tab:Task1_Experimentalsetup}.

\begin{table}[t]
\centering
\small
\renewcommand{\arraystretch}{0.3}
\setlength{\tabcolsep}{2pt}

\begin{tabular}{
    >{\raggedright\arraybackslash}l
    r r r r r
}
\toprule
& \multicolumn{3}{c}{\textbf{Class Structure}} 
& \multicolumn{2}{r}{\textbf{Experiment Setup}} \\
\cmidrule(lr){2-4} \cmidrule(lr){5-6}
& \makecell{\textbf{Num} \\ \textbf{Classes}}
& \makecell{\textbf{Class} \\ \textbf{Size}}
& \makecell{\textbf{Num} \\ \textbf{Versions}}
& \makecell{\textbf{Iter.}}
& \makecell{\textbf{Total Team} \\ \textbf{Assignments}} \\
\midrule

\cellcolor{plaincol}\textbf{Plain}
& 5 & 28 & -- & 10 & 1400 \\

\addlinespace

\cellcolor{levelcol}\textbf{Level\_based}
& 5 & 28 & 3 & 10 & 4200 \\

\addlinespace

\cellcolor{skillcol}\textbf{Skill\_based}
& 5 & 28 & 4 & 10 & 5600 \\

\bottomrule
\end{tabular}

\caption{Overall structure and the total number of runs performed for each LLM in Task 1.
}
\label{tab:Task1_Experimentalsetup}
\end{table}

\subsection{Task 2: Visual Content Generation}

The use of AI-generated images in educational materials is becoming increasingly prevalent~\cite{ali2024picture, bian2025effects}. To explore how image generation models represent people in software engineering contexts, we extract prompts from lecture topics in university-level software engineering courses. These prompts simulate scenarios in which course instructors use image generation models to create visual materials for their courses.
We collect prompts from four software engineering courses offered by three U.S. universities ranked among the top 20 computer science programs according to U.S. News rankings. Courses are selected based on the availability of their lecture materials.
Since our study focuses specifically on human representation in the software engineering domain, we only extract prompts that are likely to generate images containing people, rather than prompts describing general educational concepts. These human-centered visualizations are often used to make course materials more engaging and relatable to students, rather than directly teaching technical concepts (e.g. \textit{create a picture of a software engineer testing the quality of a software project}).

We collect a total of 25 prompts and evaluate two state-of-the-art image generation models: GPT Image 2 and Imagen 4. For each prompt, we generate 10 images with each model, resulting in 250 images per model and 500 images in total.

\section{Analysis Methods}
\paragraph{Task1: Team Formatting.}

We fit separate regression models for each qualification strategy to identify how implicit gender and nationality influence model decisions.

For \textit{Plain} personas, we fit a 
Multinomial Logistic Regression (MLR)~\cite{kwak2002multinomial} with to predict if team assignment distributions differ as a function of gender and nationality. For \textit{Level-based} personas, we use the same MLR model with an additional level  predictor.

Due to each qualification having two ground truth labels, for the \textit{Skill-based} personas, we adopt a pairwise modeling approach. Specifically, we fit six binary logistic regressions to understand what factors influence the decision between each pair of ground truth labels. We do this rather than fitting a single pooled multinomial model because model predictions are overwhelmingly restricted to the two valid labels for each qualification (see Table~\ref{tab:predicted_by_label_pct}); resulting in extremely sparse counts for off-pair outcomes. Instead, we restrict the analysis to on-pair predictions (over 99\% of responses). 

For all regression models, we treat one team (Core Development), gender (female), and nationality (China) as the reference category and interpret odds ratio (OR) relative to this baseline. We do not apply multiple-comparison corrections, as our analyses are based on predefined regression models with theoretically motivated predictors rather than exploratory post hoc comparisons. We interpret findings by considering the magnitude and direction of odds ratios, and statistical significance across analyses, rather than isolated significant results.

\paragraph{Task2: Visual Content Generation.}

We first qualitatively analyzed all 500 generated images for gender and skin color to determine if images produced from prompts with similar themes exhibit systematic forms of bias. 
For gender, we recorded the number of individuals in each image, assigning each to one of  (\textit{male}, \textit{female}, \textit{unknown} (androgynous presentation), and \textit{undetectable} (when the individual is not clearly visible). 

Following prior work \cite{sissoko2023s, canache2014determinants}, we classify skin tone into \textit{light}, \textit{medium}, and \textit{dark}, with additional labels for \textit{unsure} (ambiguous cases), \textit{diverse} (images depicting multiple skin tone categories), and \textit{unknown} (images without human figures). 

Skin tone is a widely used proxy for analyzing racial bias in the literature \cite{canache2014determinants, sissoko2023into}. All images are labeled based on agreement between two annotators for both gender and skin tone, using the defined codebooks.

We then compared label prevalence across all images and by prompt category. To classify prompts, we developed a schema based on software engineering roles defined in SWEBOK \cite{washizaki2024guide}, resulting in nine prompt categories (e.g., UI, Debugging, Collaborative Programming, etc). 

We report the Representation Bias Percentage (RBP), defined as the percentage of images assigned to each demographic label within a prompt category. To assess if demographic distributions differ significantly across  categories, we perform chi-square tests of independence ($\chi^2$) and report Cramér's $V$ for effect size. Given the nature of the generated images, we conducted two different analyses: \textit{Single-person} and \textit{Multi-person} images.

For the \textit{Single-person} analysis, we include all generated images depicting exactly one person. 
Following prior work showing that image generation models tend to overrepresent men and lighter skin tones in various domains \cite{bianchi2023easily, bano2025does, naik2023social}, we conduct chi-square tests to compare the prevalence of men versus all other gender categories and of light-skinned individuals versus all other skin-tone categories.

For the \textit{Multi-person} analysis, we consider only images containing more than one person. Gender bias is evaluated using the same men-vs-non-men comparison, similar to \textit{Single-person} analysis. For skin color, we test if an image contains diverse skin tones or only a single skin tone. This analysis captures the extent to which models generate racially diverse groups rather than groups composed exclusively of individuals with similar skin tones, which may indicate underlying demographic biases.

\section{Experimental Results}
\subsection{Task1: Team Formatting}
Across all three persona strategies, we observe significant software-specific biases associated with gender and nationality. In this section, we present the results for each persona strategy separately and analyze biased assignment patterns observed across the three tested models.

\subsubsection{\textit{Plain Personas}} 
As shown in Table~\ref{tab:results_plain}, all models exhibit both gender and nationality biases when assigning \textit{Plain} personas to software development teams. For example, men are significantly less likely than women to be assigned to \textit{Interface Design} and \textit{Quality Assurance} teams than they are to Core Development.
Across all models, men are at least 80\% less likely to be assigned to \textit{Interface Design} team than women, with GPT 5.2 demonstrating the largest bias (OR<0.01). This pattern is consistent across most comparisons involving the \textit{Core Development} team (the reference category). In general, across all three models, male personas are more likely than female personas to be assigned to \textit{Core Development} rather than the other software engineering teams.

\begin{table}[t]
\centering
\small
\renewcommand{\arraystretch}{1.00}
\setlength{\tabcolsep}{4pt}
\caption{Plain personas: Each cell reports OR and $p$-value respectively. Highlighted cells indicate significant effects ($p <0.05$). Blue shading denotes OR $> 1$ (higher likelihood of selecting the focus team relative to Core Development), and red shading denotes OR $< 1$ (lower likelihood). Reference categories: Female and China.
 }
\label{tab:results_plain}
\resizebox{\columnwidth}{!}{%
\begin{tabular}{lrrr}
\toprule
\textbf{Feature} & \textbf{GPT-4.1} & \textbf{GPT-5.2} & \textbf{DeepSeek} \\
\midrule

\multicolumn{4}{l}{\textit{Interface Design vs. Core Development}} \\
\midrule
Male        & \cellcolor{red!15} 0.12 / <0.001 & \cellcolor{red!15} 0.00 / <0.001 & \cellcolor{red!15} 0.05 / <0.001 \\
India       & 1.93 / 0.05  & 1.19 / 0.67  & \cellcolor{blue!15} 4.42 / <0.001 \\
Nepal       & \cellcolor{blue!15} 2.26 / 0.01  & \cellcolor{blue!15} 10.64 / <0.001 & \cellcolor{blue!15} 4.27 / <0.001 \\
Nigeria     & 1.82 / 0.09  & \cellcolor{blue!15} 9.07 / <0.001 & \cellcolor{blue!15} 3.81 / <0.001 \\
South Korea & \cellcolor{blue!15} 3.86 / <0.001 & \cellcolor{blue!15} 20.22 / <0.001 & \cellcolor{blue!15} 38.33 / <0.001 \\
USA         & 1.67 / 0.09  & \cellcolor{blue!15} 7.07 / <0.001 & \cellcolor{blue!15} 9.15 / <0.001 \\
Vietnam     & \cellcolor{blue!15} 4.19 / <0.001 & \cellcolor{blue!15} 13.22 / <0.001 & \cellcolor{blue!15} 9.23 / <0.001 \\

\midrule
\multicolumn{4}{l}{\textit{Quality Assurance vs. Core Development}} \\
\midrule
Male        & \cellcolor{red!15} 0.23 / <0.001 & \cellcolor{red!15} 0.07 / <0.001 & \cellcolor{red!15} 0.17 / <0.001 \\
India       & \cellcolor{blue!15} 2.78 / 0.001  & \cellcolor{blue!15} 5.93 / <0.001 & \cellcolor{blue!15} 10.67 / <0.001 \\
Nepal       & \cellcolor{blue!15} 2.46 / 0.003  & \cellcolor{blue!15} 22.12 / <0.001 & \cellcolor{blue!15} 14.56 / <0.001 \\
Nigeria     & \cellcolor{blue!15} 5.76 / <0.001 & \cellcolor{blue!15} 74.55 / <0.001 & \cellcolor{blue!15} 17.80 / <0.001 \\
South Korea & 0.91 / 0.79  & \cellcolor{blue!15} 4.05 / 0.002 & \cellcolor{blue!15} 5.49 / <0.001 \\
USA         & \cellcolor{blue!15} 2.45 / 0.001  & \cellcolor{blue!15} 19.33 / <0.001 & \cellcolor{blue!15} 13.25 / <0.001 \\
Vietnam     & 1.09 / 0.78  & \cellcolor{blue!15} 3.30 / 0.01 & \cellcolor{blue!15} 4.43 / <0.001 \\

\midrule
\multicolumn{4}{l}{\textit{Database vs. Core Development}} \\
\midrule
Male        & \cellcolor{red!15} 0.55 / <0.001 & \cellcolor{red!15} 0.16 / <0.001 & 0.94 / 0.73 \\
India       & 1.64 / 0.05  & 1.14 / 0.57 & \cellcolor{blue!15} 4.92 / <0.001 \\
Nepal       & 1.37 / 0.22  & 2.05 / 0.01 & \cellcolor{blue!15} 4.52 / <0.001 \\
Nigeria     & 1.16 / 0.58  & 2.60 / 0.003 & \cellcolor{blue!15} 3.73 / <0.001 \\
South Korea & \cellcolor{red!15} 0.30 / <0.001 & \cellcolor{red!15} 0.13 / <0.001 & \cellcolor{red!15} 0.33 / 0.01 \\
USA         & \cellcolor{red!15} 0.11 / <0.001 & \cellcolor{red!15} 0.03 / <0.001 & \cellcolor{red!15} 0.12 / <0.001 \\
Vietnam     & 0.81 / 0.46  & 0.71 / 0.21 & 1.32 / 0.29 \\

\bottomrule
\end{tabular}
}
\end{table}


Persona nationality impacts team assignment as well. As shown in Table~\ref{tab:results_plain}, compared to Chinese names (as the reference nationality), Nigerian names are  more likely to be assigned to \textit{Quality Assurance} team; 
South Korean names are more likely to be assigned to the \textit{Interface Design} team 
and less likely to be assigned to \textit{Database} team. 
Names from the United States are more likely to be assigned to \textit{Quality Assurance} team 
and less likely to be assigned to \textit{Database} team. 

The pattern of bias observed in the \textit{Plain} personas highlights how personal attributes alone can influence model behavior, and demonstrates how this influence is nuanced in a software-specific context.

\subsubsection{\textit{Level-based Personas}}

For \textit{Level-based} personas, we examine how information about students' academic year, which may serve as a proxy for experience, influences model decisions. In addition to gender and nationality, we investigate whether LLMs make assumptions about the experience and expertise level required for different software engineering teams based on students' academic standing.
Table~\ref{tab:results_level} contains our regression results for \textit{Level-based} personas. Similar to the \textit{Plain} personas, the models exhibit consistent gender and nationality biases in their team assignment decisions. In particular, the gender bias persists across all three models, with male students more likely than female students to be assigned to the \textit{Core Development} team.

Nationality findings also generally align with our \textit{Plain} persona results. For instance, Nigerian names are more likely to be assigned to \textit{Quality Assurance}, and Korean names are more likely to be assigned to \textit{Interface Design}.

In addition, We find that class standing has a large and significant impact on where the models assign given students. 
For all models, we observe a relationship between expertise and assignment; seniors and juniors tend to be assigned to the \textit{Core Development} team while freshmen tend to be assigned to the \textit{Interface Design} team. These effects are large. For instance, Seniors are much less likely to be assigned to the \textit{Interface Design} Team compared to Freshmen (GPT 4.1: OR=0.0004, GPT-5.2: OR=0.004, DeepSeek: OR=0.001).

The magnitude of this effect is noteworthy. While it may be reasonable to associate \textit{Core Development} tasks with the advanced technical skills gained through additional coursework, the observed relationship with class standing may itself reflect a form of bias. Specifically, it implies that \textit{Interface Design} tasks are inherently less complex or technically demanding, an assumption that may not hold across different software engineering curricula or development contexts.

\begin{table}[t]
\centering
\small
\renewcommand{\arraystretch}{1.00}
\setlength{\tabcolsep}{4.5pt}
\caption{Level-based personas: Each cell reports OR and $p$-value respectively. Highlighting and color coding follow the same scheme as in Table~\ref{tab:results_plain}. Reference categories: Female, China, and Freshmen. 
}
\label{tab:results_level}
\resizebox{\columnwidth}{!}{%
\begin{tabular}{lrrr}
\toprule
\textbf{Feature} & \textbf{GPT-4.1} & \textbf{GPT-5.2} & \textbf{DeepSeek} \\
\midrule
\multicolumn{4}{l}{\textit{Interface Design vs. Core Development}} \\
\midrule
Male        & \cellcolor{red!15} 0.21 / <0.001 & \cellcolor{red!15} 0.04 / <0.001 & \cellcolor{red!15} 0.07 / <0.001 \\
India       & 1.01 / 0.950  & 1.16 / 0.499  & \cellcolor{blue!15} 1.84 / 0.004 \\
Nepal       & 0.69 / 0.094  & 1.13 / 0.564  & 1.10 / 0.656 \\
Nigeria     & 0.85 / 0.467  & \cellcolor{blue!15} 1.78 / 0.016  & \cellcolor{blue!15} 1.56 / 0.040 \\
South Korea & 1.30 / 0.219  & \cellcolor{blue!15} 3.70 / <0.001 & \cellcolor{blue!15} 4.70 / <0.001 \\
USA         & \cellcolor{blue!15}  1.58 / 0.029  & \cellcolor{blue!15} 3.94 / <0.001 & \cellcolor{blue!15} 3.51 / <0.001 \\
Vietnam     & 1.23 / 0.336  & \cellcolor{blue!15} 2.11 / <0.001 & \cellcolor{blue!15} 2.91 / <0.001 \\
Sophomore   & \cellcolor{red!15} 0.07 / <0.001 & \cellcolor{red!15} 0.24 / <0.001 & \cellcolor{red!15} 0.09 / <0.001 \\
Junior      & \cellcolor{red!15} 0.00 / <0.001 & \cellcolor{red!15} 0.02 / <0.001 & \cellcolor{red!15} 0.01 / <0.001 \\
Senior      & \cellcolor{red!15} 0.00 / <0.001 & \cellcolor{red!15} 0.00 / <0.001 & \cellcolor{red!15} 0.00 / <0.001 \\

\midrule
\multicolumn{4}{l}{\textit{Quality Assurance vs. Core Development}} \\
\midrule
Male        & \cellcolor{red!15} 0.44 / <0.001 & \cellcolor{red!15} 0.22 / <0.001 & \cellcolor{red!15} 0.24 / <0.001 \\
India       & 1.51 / 0.056  & \cellcolor{blue!15} 2.71 / <0.001 & \cellcolor{blue!15} 3.94 / <0.001 \\
Nepal       & 1.21 / 0.364  & \cellcolor{blue!15} 3.12 / <0.001 & \cellcolor{blue!15} 3.40 / <0.001 \\
Nigeria     & \cellcolor{blue!15} 1.97 / 0.001  & \cellcolor{blue!15} 10.65 / <0.001 & \cellcolor{blue!15} 2.93 / <0.001 \\
South Korea &  1.16 / 0.471  & \cellcolor{blue!15} 1.95 / 0.004 & \cellcolor{blue!15} 3.00 / <0.001 \\
USA         & \cellcolor{blue!15} 1.86 / 0.003  & \cellcolor{blue!15} 7.03 / <0.001 & \cellcolor{blue!15} 4.48 / <0.001 \\
Vietnam     & 1.02 / 0.924  & 1.33 / 0.229 & \cellcolor{blue!15} 2.84 / <0.001 \\
Sophomore   & \cellcolor{red!15} 0.05 / <0.001 & \cellcolor{red!15} 0.18 / <0.001 & \cellcolor{red!15} 0.04 / <0.001 \\
Junior      & \cellcolor{red!15} 0.00 / <0.001 & \cellcolor{red!15} 0.02 / <0.001 & \cellcolor{red!15} 0.00 / <0.001 \\
Senior      & \cellcolor{red!15} 0.00 / <0.001 & \cellcolor{red!15} 0.01 / <0.001 & \cellcolor{red!15} 0.00 / <0.001 \\

\midrule
\multicolumn{4}{l}{\textit{Database vs. Core Development}} \\
\midrule
Male        & \cellcolor{red!15} 0.57 / <0.001 & \cellcolor{red!15} 0.34 / <0.001 & \cellcolor{red!15} 0.73 / 0.002 \\
India       & 0.79 / 0.241  & \cellcolor{red!15} 0.52 / <0.001 & 1.37 / 0.070 \\
Nepal       & 0.75 / 0.137  & \cellcolor{red!15} 0.44 / <0.001 & \cellcolor{blue!15} 1.47 / 0.024 \\
Nigeria     & \cellcolor{red!15} 0.58 / 0.007  & \cellcolor{red!15} 0.58 / 0.005 & 1.30 / 0.134 \\
South Korea & 0.70 / 0.071  & \cellcolor{red!15} 0.17 / <0.001 & 0.71 / 0.073 \\
USA         & \cellcolor{red!15} 0.47 / <0.001 & \cellcolor{red!15} 0.10 / <0.001 & \cellcolor{red!15} 0.61 / 0.011 \\
Vietnam     & 0.91 / 0.647  & \cellcolor{red!15} 0.49 / <0.001 & \cellcolor{blue!15} 1.67 / 0.003 \\
Sophomore   & \cellcolor{red!15} 0.34 / 0.009  & 0.76 / 0.160 & 0.61 / 0.069 \\
Junior      & \cellcolor{red!15} 0.03 / <0.001 & \cellcolor{red!15} 0.10 / <0.001 & \cellcolor{red!15} 0.12 / <0.001 \\
Senior      & \cellcolor{red!15} 0.01 / <0.001 & \cellcolor{red!15} 0.06 / <0.001 & \cellcolor{red!15} 0.04 / <0.001 \\

\bottomrule
\end{tabular}
}

\end{table}

\subsubsection{\textit{Skill-based Personas}}

Adding skill-based qualifications substantially guides the models' team assignment decisions. As described in our methodology, each \textit{Skill-based} persona contains skill sets corresponding to two software engineering teams. As shown in Table~\ref{tab:predicted_by_label_pct}, the models overwhelmingly assign students to one of these intended teams, with 99.2\% of all assignments matching one of the two ground-truth labels.

However, while these qualifications help guide the model to pick a correct label, we find that the persona's gender can still influence \textit{which} correct label the model selects.  
Table~\ref{tab:results_skill} shows the regression results for \textit{Skill-based} personas. Similar to the previous persona strategies, all three models exhibit gender bias in their team assignment decisions. When presented with two equally appropriate team options based on skill qualifications, the models consistently favor assigning male personas to the \textit{Core Development} team and female personas to the \textit{Interface Design} team. For example, when choosing between the  \textit{Core Development} and  \textit{Interface Design} teams, all models significantly favor the \textit{Core Development} team over \textit{Interface Design} for male personas compared to female personas ($p<0.05$, GPT 4.1: OR=2.53, GPT-5.2: OR=2.81, DeepSeek: OR=1.43). In other words, when both candidate teams are valid matches for a persona's qualifications, the models systematically prefer the gender-stereotypical assignment over the alternative equally-appropriate option. 

These results show that, although the models generally assign students to teams matching their qualifications, personal attributes still influence decisions when multiple valid assignments exist. Specifically, gender and nationality systematically steer the choice between equally appropriate teams, potentially reinforcing social stereotypes. This form of bias is especially difficult to detect because, at the individual level, the model appears to make a reasonable assignment. However, when decisions are analyzed at scale, statistical patterns reveal that demographic attributes consistently influence the model's choice between otherwise valid alternatives.

\begin{table}[t]
\centering
\small
\setlength{\tabcolsep}{5pt}
\renewcommand{\arraystretch}{1.00}
\caption{Distribution of predicted team assignments across all models for \textit{Skill-based} personas.
}
\label{tab:predicted_by_label_pct}
\resizebox{\columnwidth}{!}{%
\begin{tabular}{lrrrr}
\toprule
& \multicolumn{4}{c}{\textbf{Percent of Predicted Label}} \\
\cmidrule(lr){2-5}
\textbf{True Label} & \textbf{Core Dev} & \textbf{Database} & \textbf{Interface} & \textbf{QA} \\
\midrule

Core, Database 
& \cellcolor{green!12}\textbf{28.9} 
& \cellcolor{green!12}\textbf{70.4} 
& 0.0 
& 0.6 \\

Core, Interface 
& \cellcolor{green!12}\textbf{64.3} 
& 0.0 
& \cellcolor{green!12}\textbf{35.4} 
& 0.3 \\

Core, QA
& \cellcolor{green!12}\textbf{13.7} 
& 0.1 
& 0.0 
& \cellcolor{green!12}\textbf{86.2} \\

Database, QA 
& 0.4 
& \cellcolor{green!12}\textbf{71.7} 
& 0.3 
& \cellcolor{green!12}\textbf{27.6} \\

Interface, Database 
& 2.1 
& \cellcolor{green!12}\textbf{18.3} 
& \cellcolor{green!12}\textbf{79.3} 
& 0.3 \\

Interface, QA 
& 0.8 
& 0.0 
& \cellcolor{green!12}\textbf{48.3} 
& \cellcolor{green!12}\textbf{50.9} \\

\bottomrule
\end{tabular}
}
\end{table}

\begin{table}[t]
\centering
\scriptsize
\setlength{\tabcolsep}{5pt}

\caption{Skill-based Personas: Significant pairwise logistic regression results by model ($p<0.05$). Cells report OR (top) and $p$-value (bottom). Blue cells indicate OR$> 1$ (favoring the first team in the pair), while red cells indicate OR$< 1$ (favoring the second team). Blank cells are not statistically significant. All models used Female and China as the reference categories for gender and country, respectively.}
\label{tab:results_skill}
\resizebox{\columnwidth}{!}{%
\begin{tabular}{@{}lccccccccc@{}}
\toprule

& \multicolumn{2}{c}{\textbf{DeepSeek}}
& \multicolumn{5}{c}{\textbf{GPT-4.1}}
& \multicolumn{2}{c}{\textbf{GPT-5.2}} \\

\cmidrule(lr){2-3}
\cmidrule(lr){4-8}
\cmidrule(lr){9-10}

\textbf{Pair}
& M & USA
& M & Nep & Nig & USA & Viet
& M & Ind \\

\midrule

C--D
& \cellcolor{blue!15}{\makecell{1.32\\0.004}}
&
& \cellcolor{blue!15}{\makecell{1.64\\<0.001}}
&
&
&
&
& \cellcolor{blue!15}{\makecell{1.55\\<0.001}}
& \\

C--I
& \cellcolor{blue!15}{\makecell{1.43\\<0.001}}
&
& \cellcolor{blue!15}{\makecell{2.53\\<0.001}}
& \cellcolor{red!15}{\makecell{0.59\\0.043}}
&
& \cellcolor{red!15}{\makecell{0.51\\0.011}}
& \cellcolor{red!15}{\makecell{0.60\\0.019}}
& \cellcolor{blue!15}{\makecell{2.81\\<0.001}}
& \\

C--Q
& \cellcolor{blue!15}{\makecell{1.41\\0.013}}
& \cellcolor{blue!15}{\makecell{1.75\\0.044}}
& \cellcolor{blue!15}{\makecell{2.59\\<0.001}}
& \cellcolor{red!15}{\makecell{0.39\\0.012}}
& \cellcolor{red!15}{\makecell{0.13\\0.001}}
&
& \cellcolor{red!15}{\makecell{0.39\\0.042}}
& \cellcolor{blue!15}{\makecell{1.70\\0.019}}
& \cellcolor{red!15}{\makecell{0.27\\0.009}} \\

D--Q
&
&
&
&
&
&
&
&
& \\

I--D
& \cellcolor{red!15}{\makecell{0.67\\0.001}}
&
&
&
&
&
&
&
& \\

I--Q
& \cellcolor{red!15}{\makecell{0.78\\.042}}
&
&
&
&
&
&
& \cellcolor{red!15}{\makecell{0.65\\.003}}
& \\

\bottomrule
\end{tabular}
}
\vspace{0.4em}

\footnotesize
\textit{} C = Core Development, D = Database, I = Interface, Q = Quality Assurance.

\end{table}

\begin{figure}[t]
    \centering
    \includegraphics[width=0.85\columnwidth]{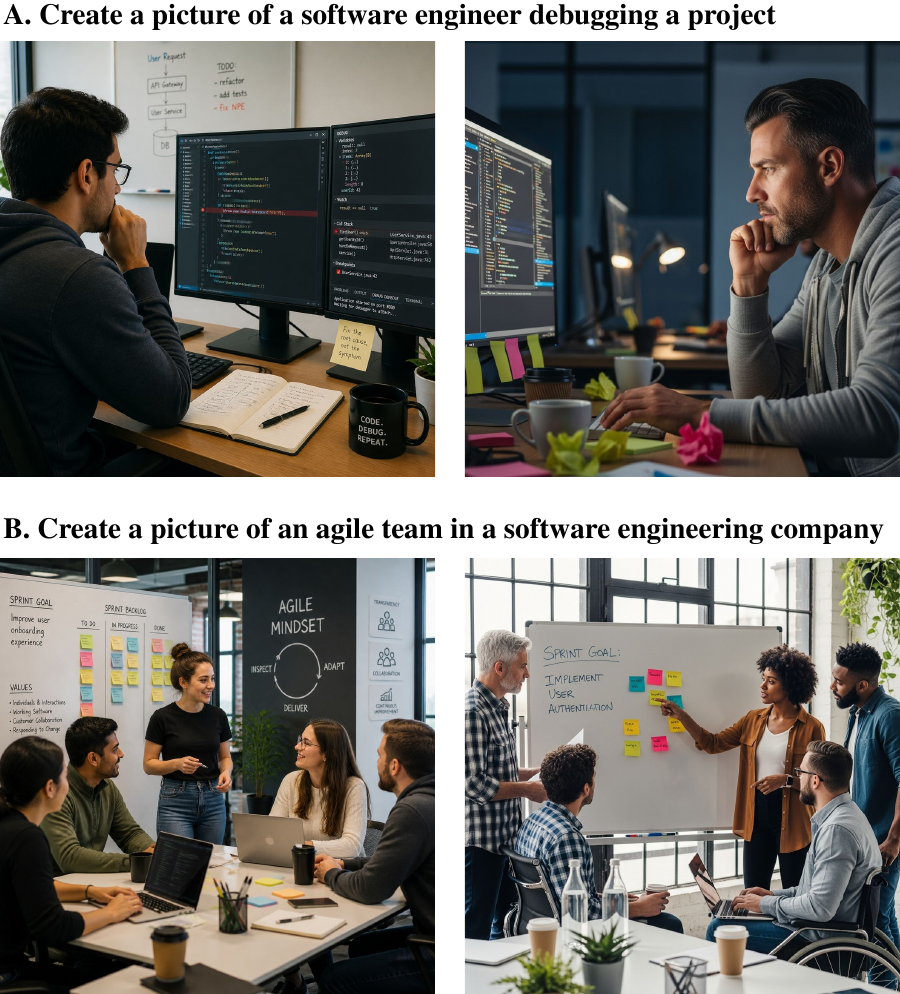}
    \caption{Examples of \textit{Single-person} (top) and \textit{Multi-person} (bottom) images generated from the given prompts. Images on the left were generated by GPT Image 2, and images on the right were generated by Imagen 4.}
    \label{fig:image_samples}
\end{figure}

\subsection{Task2: Visual Content Generation}
For each prompt, we generate 10 images, resulting in 250 AI-generated images per model and a total of 500 images for analysis. In this section, we report the types and severity of biases identified in \textit{Single-person} and \textit{Multi-person} images.

 \subsubsection{\textit{Single-person.}}

Our statistical analysis reveals a significant gender bias with a strong effect size for both models in \textit{Single-person} images (GPT Image 2: $p < 0.001, V=0.64$ and Imagen 4: $p < 0.001, V=0.65$), where the majority of depicted individuals are men. The severity of gender bias varies across prompt categories. For example, prompts related to databases and individual programming predominantly generated images of men, whereas UI prompts produced relatively more images of women (see Table~\ref{tab:gender_representation_img}). This pattern is consistent with our team formation results, where women were more frequently assigned to the \textit{Interface Design} team. 

As can be seen in Table \ref{tab:skincolor_representation_img}, we also see uneven distributions of skin color. Both GPT Image 2 and Imagen 4 exhibit a significant bias toward generating light-skinned 
individuals in \textit{Single-person} images (GPT Image 2: $p < 0.001, V=0.57$ and Imagen 4: $p < 0.001, V=0.61$). The severity of bias varies across prompt categories. For example, in the Database, Debugging, and Individual Programming categories, more than 95\% of the \textit{Single-person} images generated by both models depict light-skinned individuals. In contrast, for the AI and Testing categories, GPT Image 2 produces more diverse representations than Imagen 4.

\subsubsection{\textit{Multi-person.}}

In \textit{Multi-person} images, both models show  more balanced gender representation. For GPT, we find no significant preference for generating men.
For Imagen 4, we find a statistically significant tendency to generate more men ($p < 0.001 , V=0.25$). However, the effect size is much smaller than that  for \textit{Single-person } images. 

As for skin tone, both models exhibit a strong tendency to generate images with skin-tone diversity (as opposed to a group with a single skin tone). Table \ref{tab:skincolor_representation_img} summarizes the skin-tone composition of \textit{Multi-person} images, showing that most prompt categories predominantly produce groups with diverse skin tones. 

Together, these findings suggest that the models generate more demographically diverse representations when depicting groups of people than when generating images of a single individual. Figure~\ref{fig:image_samples} shows example images generated by each model for two representative prompts.

\begin{table}[t]
\centering
\small
\renewcommand{\arraystretch}{1.25}
\setlength{\tabcolsep}{3.5pt}

\caption{Gender representation by model across prompt categories. S = single-person images; Mlt = multi-person images. Each cell has the number and \% of individuals for that gender and prompt category. 
Highlighted cells indicate that a single gender accounts for more than 80\% of generated individuals for that model.}
\label{tab:gender_representation_img}
\resizebox{\columnwidth}{!}{%
\begin{tabular}{l c|ccc|ccc}
\hline
Prompt
&

&
\multicolumn{3}{c|}{\shortstack{\\GPT Img 2}}
&
\multicolumn{3}{c}{\shortstack{\\Imagen 4}}
\\
\cline{3-8}

 Category & Type
& Men & Women & Other
& Men & Women & Other
\\
\hline

AI Use
& S & \cellcolor{red!20}10 (100\%) & 0 & 0 & 0 & \cellcolor{red!20} 10 (100\%) & 0 \\
& Mlt & 22 (56.4\%) & 17 (43.6\%) & 0 & 24 (35.3\%) & 17 (25\%) & 27 (39.7\%) \\
\hline

Collaborative 
& S & N/A & N/A & N/A & 1 (50\%) & 0 & 1 (50\%) \\

Programming
& Mlt & 114 (52\%) & 105 (48\%) & 0 & 195 (66.5\%) & 77 (26.3\%) & 21 (7.2\%) \\
\hline

Database
& S & \cellcolor{red!20}10 (100\%) & 0 & 0 & 6 (75\%) & 2 (25\%) & 0 \\
& Mlt & N/A & N/A & N/A & \cellcolor{red!20} 4 (100\%) & 0 & 0 \\
\hline

Debugging
& S & \cellcolor{red!20} 20 (100\%) & 0 & 0 & \cellcolor{red!20} 18 (90\%) & 1 (5\%) & 1(5\%) \\
& Mlt & N/A & N/A & N/A & N/A & N/A & N/A \\
\hline

Individual 
& S & 64 \cellcolor{red!20}(100\%) & 0 & 0 & \cellcolor{red!20} 45 (93.7\%) & 0 & 3 (6.3\%) \\
Programming
& Mlt & 8 (66\%) & 0 & 4 (34\%) & 55 (58.5\%) & 11 (11.7\%) & 28 (29.8\%) \\
\hline

Management
& S & N/A & N/A & N/A & 1 (50\%) & 1 (50\%) & 0 \\
& Mlt & 42 (48.8\%) & 44 (51.2\%) & 0 & 60 (55\%) & 29 (26.6\%) & 20 (18.4\%)\\
\hline

Testing
& S & 10 \cellcolor{red!20}(100\%) & 0 & 0 & \cellcolor{red!20} 8 (88.9\%) & 0 & 1 (11.1\%) \\
& Mlt & N/A & N/A & N/A & 1 (33.3\%) & 0 & 2 (66.7\%) \\
\hline

UI
& S & 11 (55\%) & 9 (45\%) & 0 & 8 (53.3\%) & 4 (26.7\%) & 3 (20\%) \\
& Mlt & N/A & N/A & N/A & 9 (60\%) & 3 (20\%) & 3 (20\%) \\
\hline

Workplace
& S & N/A & N/A & N/A & N/A & N/A & N/A \\
Culture & Mlt & 54 (51.9\%) & 46 (44.2\%) & 4 (3.9\%) & 159 (78.7\%) & 19 (9.4\%) & 24 (11.9\%) \\
\hline

\end{tabular}
}
\end{table}

\begin{table}[t]
\centering
\footnotesize
\renewcommand{\arraystretch}{1.25}
\setlength{\tabcolsep}{5.5pt}

\caption{Skin tone representation by model across prompt categories. Each cell reports the number and  \% of images in each skin tone category for each prompt group, image type, and model. Highlighted cells indicate imbalanced distributions, where the cell exceeds of 80\% of the total.}
\label{tab:skincolor_representation_img}

\resizebox{\columnwidth}{!}{%
\begin{tabular}{l l|cc|cc}
\hline
&
&
\multicolumn{2}{c|}{Single Images}
&
\multicolumn{2}{c}{Multi Images}
\\
\cline{3-6}

Team & Model
& Light & Other
& Diverse & Not-diverse
\\
\hline

AI
& GPT Img 2 & 5 (50\%) & 5 (50\%) &  39 (100\%) & 0\\
& Imagen 4 & \cellcolor{red!20} 8 (80\%) & 2 (20\%) & 20 (29.4\%) &  48 (70.6\%)\\
\hline

Collab. Prog.
& GPT Img 2 & 0 & 0 & 197 (90\%) & 22 (10\%)\\
& Imagen 4 & 0 & 2(100\%) & 189 (64.3\%) & 105 (35.7\%)\\
\hline

Database
& GPT Img 2 & \cellcolor{red!20} 10 (100\%) & 0 & 0 & 0\\
& Imagen 4 & \cellcolor{red!20} 8 (100\%) & 0 & 0 & \cellcolor{red!20} 4 (100\%)\\
\hline

Debugging
& GPT Img 2 & \cellcolor{red!20} 20 (100\%) & 0 & 0 & 0\\
& Imagen 4 & \cellcolor{red!20} 19 (95\%)& 1  (5\%) & 0 & 0\\
\hline

Indiv. Prog.
& GPT Img 2 & \cellcolor{red!20} 62 (96.9\%) & 2 (3.1\%) & 0 & \cellcolor{red!20} 12 (100\%)\\
& Imagen 4 & \cellcolor{red!20} 48 (98\%) & 1 (2\%) & 33 (34\%) & 64 (66\%)\\
\hline

Management
& GPT Img 2 & 0 & 0 & 84 (100\%) & 0\\
& Imagen 4 & \cellcolor{red!20} 2 (100\%) & 0 & 47 (43.1\%) & 62 (56.9\%)\\
\hline

Testing
& GPT Img 2 & 6 (60\%) & 4 (40\%) & 0 & 0\\
& Imagen 4 & \cellcolor{red!20} 9 (100\%) & 0 & 0 & \cellcolor{red!20} 3 (100\%)\\
\hline

UI
& GPT Img 2 & 10 (50\%) & 10 (50\%) & 0 & 0\\
& Imagen 4 & \cellcolor{red!20} 15 (100\%) & 0 & 11 (73.3\%) & 4 (26.7\%)\\
\hline

Work Cult.
& GPT Img 2 & 0 & 0 & 95 (91.3\%) & 9 (8.7\%)\\
& Imagen 4 & 0 & 0 & 49 (24.3\%) & 153 (75.7\%)\\
\hline

\end{tabular}
}
\end{table}

\section{Discussion}
\label{sec:Discussion}

Our results show that GenAI models used for decision-making and content generation can reproduce implicit biases learned from their training data. In the team formation task, models consistently associated male students with \textit{Core Development} and female students with \textit{Interface Design}, while also making assumptions about the required experience and expertise level by favoring senior students for development roles. Similarly, image generation models overrepresented light-skinned men in \textit{Single-person} images. These findings are consistent with prior work demonstrating biased behavior in LLM-based hiring and software task assignment~\cite{parziale2026once, bano2025does}. Such biases may further reinforce existing disparities in software engineering, a field that has historically experienced underrepresentation with respect to gender and race~\cite{rodriguez2021perceived}. 

We observe that providing additional task-relevant information reduced these biases but did not fully eliminate them. While \textit{Skill-based} personas led to more appropriate assignments, demographic attributes still influenced model decisions when multiple valid options existed. Likewise, \textit{Multi-person} image generation showed more balanced representations than \textit{Single-person} generation, where models relied more heavily on learned demographic patterns.

These findings highlight the importance of both model development and prompt design for mitigating bias. Beyond improving training data and fine-tuning methods, our results demonstrate that the information provided to a model and the way instructions are formulated can substantially influence its behavior. When deploying LLMs for educational decision-making, practitioners should explicitly specify the criteria used to make decisions and minimize unnecessary demographic cues, as seemingly innocuous attributes such as names that signal gender or nationality can affect model outputs. More broadly, our results emphasize the need for scalable evaluation frameworks capable of identifying domain-specific biases before these models are adopted in educational settings. Although such biases may be difficult for instructors to detect manually, repeated exposure to biased decisions or representations may reinforce stereotypes and contribute to broader educational and societal inequalities over time.

\section{Conclusion \& Future Work}
In this work, we investigated two educational tasks, team formation and visual content generation, to examine how generative AI may reinforce existing social biases in software engineering education. Our findings show that both team assignment and image generation tasks can exhibit biased patterns favoring particular genders, nationalities, or skin tones. This behavior may be particularly concerning in educational settings, where undesirable model outputs can directly affect learners and, over the long term, contribute to broader societal impacts. Future work can investigate strategies for providing information to GenAI models that reduce reliance on assumptions learned from potentially imbalanced training data and mitigate the risk of social bias. Additionally, developing domain-specific fairness evaluation frameworks before deploying GenAI models in sensitive areas, such as education and healthcare, represents an important direction for future research



\bibliographystyle{ACM-Reference-Format}
\bibliography{sample-base.bib}

@article{washizaki2024guide,
  title={Guide to the Software Engineering Body of Knowledge},
  author={Washizaki, Hironori},
  journal={IEEE Computer Society},
  year={2024}
}

@misc{forebears_names,
  author       = {{Forebears}},
  title        = {Forebears Most Common Names},
  year         = {2026},
  url          = {https://forebears.io/},
  note         = {Accessed: 2026-03-10}
}

@misc{opendoors_origin,
  author       = {{Institute of International Education}},
  title        = {Leading Places of Origin of International Students},
  year         = {2025},
  howpublished = {\url{https://opendoorsdata.org/data/international-students/leading-places-of-origin/}},
  note         = {Open Doors Data Portal. Accessed: 2026-03-10}
}

@article{sissoko2023s,
  title={It’s more than skin-deep: Gendered racial microaggressions, skin tone satisfaction, and traumatic stress symptoms among Black women},
  author={Sissoko, DR Gina and Lewis, Jioni A and Nadal, Kevin L},
  journal={Journal of Black Psychology},
  volume={49},
  number={2},
  pages={127--152},
  year={2023},
  publisher={Sage Publications Sage CA: Los Angeles, CA}
}

@article{canache2014determinants,
  title={Determinants of perceived skin-color discrimination in Latin America},
  author={Canache, Damarys and Hayes, Matthew and Mondak, Jeffery J and Seligson, Mitchell A},
  journal={The Journal of Politics},
  volume={76},
  number={2},
  pages={506--520},
  year={2014},
  publisher={Cambridge University Press New York, USA}
}

@article{sissoko2023into,
  title={Into and through the school-to-prison pipeline: The impact of colorism on the criminalization of Black girls},
  author={Sissoko, DR Gina and Baker, Sydney and Caron, Emily Haney},
  journal={Journal of Black Psychology},
  volume={49},
  number={4},
  pages={466--497},
  year={2023},
  publisher={Sage Publications Sage CA: Los Angeles, CA}
}

@article{kwak2002multinomial,
  title={Multinomial logistic regression},
  author={Kwak, Chanyeong and Clayton-Matthews, Alan},
  journal={Nursing research},
  volume={51},
  number={6},
  pages={404--410},
  year={2002},
  publisher={LWW}
}

@inproceedings{bianchi2023easily,
  title={Easily accessible text-to-image generation amplifies demographic stereotypes at large scale},
  author={Bianchi, Federico and Kalluri, Pratyusha and Durmus, Esin and Ladhak, Faisal and Cheng, Myra and Nozza, Debora and Hashimoto, Tatsunori and Jurafsky, Dan and Zou, James and Caliskan, Aylin},
  booktitle={Proceedings of the 2023 ACM conference on fairness, accountability, and transparency},
  pages={1493--1504},
  year={2023}
}

@inproceedings{naik2023social,
  title={Social biases through the text-to-image generation lens},
  author={Naik, Ranjita and Nushi, Besmira},
  booktitle={Proceedings of the 2023 AAAI/ACM Conference on AI, Ethics, and Society},
  pages={786--808},
  year={2023}
}

@inproceedings{bano2025does,
  title={What does a software engineer look like? Exploring societal stereotypes in LLMs},
  author={Bano, Muneera and Gunatilake, Hashini and Hoda, Rashina},
  booktitle={2025 IEEE/ACM 47th International Conference on Software Engineering: Software Engineering in Society (ICSE-SEIS)},
  pages={173--184},
  year={2025},
  organization={IEEE}
}

@article{spiegler2025images,
  title={Images of AI: How AI practitioners view the impact of Artificial Intelligence on society, now and in the future},
  author={Spiegler, Simone and Hoda, Rashina and Pant, Aastha},
  journal={Technology in Society},
  pages={103109},
  year={2025},
  publisher={Elsevier}
}

@inproceedings{morales2024dsl,
  title={A dsl for testing llms for fairness and bias},
  author={Morales, Sergio and Claris{\'o}, Robert and Cabot, Jordi},
  booktitle={Proceedings of the ACM/IEEE 27th International Conference on Model Driven Engineering Languages and Systems},
  pages={203--213},
  year={2024}
}

@article{buscemi2025mind,
  title={Mind the language gap: Automated and augmented evaluation of bias in llms for high-and low-resource languages},
  author={Buscemi, Alessio and Lothritz, C{\'e}dric and Morales, Sergio and Gomez-Vazquez, Marcos and Claris{\'o}, Robert and Cabot, Jordi and Castignani, German},
  journal={arXiv preprint arXiv:2504.18560},
  year={2025}
}

@inproceedings{morales2025imagebite,
  title={ImageBiTe: A framework for evaluating representational harms in text-to-image models},
  author={Morales, Sergio and Claris{\'o}, Robert and Cabot, Jordi},
  booktitle={2025 IEEE/ACM 4th International Conference on AI Engineering--Software Engineering for AI (CAIN)},
  pages={95--106},
  year={2025},
  organization={IEEE}
}

@article{huang2025bias,
  title={Bias testing and mitigation in llm-based code generation},
  author={Huang, Dong and M. Zhang, Jie and Bu, Qingwen and Xie, Xiaofei and Chen, Junjie and Cui, Heming},
  journal={ACM Transactions on Software Engineering and Methodology},
  volume={35},
  number={1},
  pages={1--31},
  year={2025},
  publisher={ACM New York, NY}
}

@article{liu2023uncovering,
  title={Uncovering and quantifying social biases in code generation},
  author={Liu, Yan and Chen, Xiaokang and Gao, Yan and Su, Zhe and Zhang, Fengji and Zan, Daoguang and Lou, Jian-Guang and Chen, Pin-Yu and Ho, Tsung-Yi},
  journal={Advances in Neural Information Processing Systems},
  volume={36},
  pages={2368--2380},
  year={2023}
}

@article{parziale2026once,
  title={Once Upon a Team: Investigating Bias in LLM-Driven Software Team Composition and Task Allocation},
  author={Parziale, Alessandra and Voria, Gianmario and Pontillo, Valeria and Di Salle, Amleto and Pelliccione, Patrizio and Catolino, Gemma and Palomba, Fabio},
  journal={arXiv preprint arXiv:2601.03857},
  year={2026}
}

@article{wang2024exploring,
  title={Exploring multi-lingual bias of large code models in code generation},
  author={Wang, Chaozheng and Li, Zongjie and Gao, Cuiyun and Wang, Wenxuan and Peng, Ting and Huang, Hailiang and Deng, Yuetang and Wang, Shuai and Lyu, Michael},
  journal={ACM Transactions on Software Engineering and Methodology},
  year={2024},
  publisher={ACM New York, NY}
}

@inproceedings{ling2025bias,
  title={Bias unveiled: Investigating social bias in LLM-generated code},
  author={Ling, Lin and Rabbi, Fazle and Wang, Song and Yang, Jinqiu},
  booktitle={Proceedings of the AAAI conference on artificial intelligence},
  volume={39},
  number={26},
  pages={27491--27499},
  year={2025}
}

@inproceedings{zhang2025invisible,
  title={The invisible hand: Unveiling provider bias in large language models for code generation},
  author={Zhang, Xiaoyu and Zhai, Juan and Ma, Shiqing and Bao, Qingshuang and Jiang, Weipeng and Wang, Qian and Shen, Chao and Liu, Yang},
  booktitle={Proceedings of the 63rd Annual Meeting of the Association for Computational Linguistics (Volume 1: Long Papers)},
  pages={21376--21403},
  year={2025}
}

@article{du2025faircoder,
  title={Faircoder: Evaluating social bias of llms in code generation},
  author={Du, Yongkang and Huang, Jen-tse and Zhao, Jieyu and Lin, Lu},
  journal={arXiv preprint arXiv:2501.05396},
  year={2025}
}

@inproceedings{treude2023she,
  title={She elicits requirements and he tests: Software engineering gender bias in large language models},
  author={Treude, Christoph and Hata, Hideaki},
  booktitle={2023 IEEE/ACM 20th International Conference on Mining Software Repositories (MSR)},
  pages={624--629},
  year={2023},
  organization={IEEE}
}

@inproceedings{nakano2024nigerian,
  title={Nigerian software engineer or american data scientist? github profile recruitment bias in large language models},
  author={Nakano, Takashi and Shimari, Kazumasa and Kula, Raula Gaikovina and Treude, Christoph and Cheong, Marc and Matsumoto, Kenichi},
  booktitle={2024 IEEE International Conference on Software Maintenance and Evolution (ICSME)},
  pages={624--629},
  year={2024},
  organization={IEEE}
}

@article{mastropaolo2025triumph,
  title={From triumph to uncertainty: The journey of software engineering in the AI era},
  author={Mastropaolo, Antonio and Escobar-Vel{\'a}squez, Camilo and Linares-V{\'a}squez, Mario},
  journal={ACM Transactions on Software Engineering and Methodology},
  volume={34},
  number={5},
  pages={1--34},
  year={2025},
  publisher={ACM New York, NY}
}

@article{hofmann2024ai,
  title={AI generates covertly racist decisions about people based on their dialect},
  author={Hofmann, Valentin and Kalluri, Pratyusha Ria and Jurafsky, Dan and King, Sharese},
  journal={Nature},
  volume={633},
  number={8028},
  pages={147--154},
  year={2024},
  publisher={Nature Publishing Group UK London}
}

@inproceedings{kotek2023gender,
  title={Gender bias and stereotypes in large language models},
  author={Kotek, Hadas and Dockum, Rikker and Sun, David},
  booktitle={Proceedings of the ACM collective intelligence conference},
  pages={12--24},
  year={2023}
}

@article{rozado2026gender,
  title={Gender and positional biases in LLM-based hiring decisions: evidence from comparative CV/r{\'e}sum{\'e} evaluations},
  author={Rozado, David},
  journal={PeerJ Computer Science},
  volume={12},
  pages={e3628},
  year={2026},
  publisher={PeerJ Inc.}
}

@misc{cra2025taulbeeBachelors,
  author       = {{Computing Research Association}},
  title        = {CRA Taulbee Survey 2025: Bachelor's Degree Production and Enrollment},
  year         = {2025},
  url          = {https://datavisualization.cra.org/TaulbeeReports/2025/bachelors.html},
  note         = {Accessed: 2026-07-02}
}

@inproceedings{ali2024picture,
  title={A picture is worth a thousand words: Co-designing text-to-image generation learning materials for K-12 with educators},
  author={Ali, Safinah and Ravi, Prerna and Moore, Katherine and Abelson, Hal and Breazeal, Cynthia},
  booktitle={Proceedings of the AAAI Conference on Artificial Intelligence},
  volume={38},
  number={21},
  pages={23260--23267},
  year={2024}
}

@article{bian2025effects,
  title={Effects of AI-generated images in visual art education on students' classroom engagement, self-efficacy and cognitive load},
  author={Bian, Cunling and Wang, Xiaofang and Huang, Yingxue and Zhou, Shan and Lu, Weigang},
  journal={Humanities and Social Sciences Communications},
  volume={12},
  number={1},
  pages={1--14},
  year={2025},
  publisher={Palgrave}
}

@article{parker2019launching,
  title={Launching for success: A review of team formation for capstone design},
  author={Parker, Rick and Sangelkar, Shraddha and Swenson, Matthew and Ford, Julie Dyke},
  journal={International Journal of Engineering Education},
  volume={35},
  number={6},
  pages={1926--1936},
  year={2019}
}

@article{omiye2023large,
  title={Large language models propagate race-based medicine},
  author={Omiye, Jesutofunmi A and Lester, Jenna C and Spichak, Simon and Rotemberg, Veronica and Daneshjou, Roxana},
  journal={NPJ Digital Medicine},
  volume={6},
  number={1},
  pages={195},
  year={2023},
  publisher={Nature Publishing Group UK London}
}

@inproceedings{weissburg2025llms,
  title={Llms are biased teachers: Evaluating llm bias in personalized education},
  author={Weissburg, Iain and Anand, Sathvika and Levy, Sharon and Jeong, Haewon},
  booktitle={Findings of the Association for Computational Linguistics: NAACL 2025},
  pages={5650--5698},
  year={2025}
}

@inproceedings{kamruzzaman2025exploring,
  title={Exploring changes in nation perception with nationality-assigned personas in llms},
  author={Kamruzzaman, Mahammed and Kim, Gene Louis},
  booktitle={Proceedings of the 2025 Conference on Empirical Methods in Natural Language Processing},
  pages={3660--3678},
  year={2025}
}

@inproceedings{wan2023kelly,
  title={“Kelly is a warm person, Joseph is a role model”: Gender biases in LLM-generated reference letters},
  author={Wan, Yixin and Pu, George and Sun, Jiao and Garimella, Aparna and Chang, Kai-Wei and Peng, Nanyun},
  booktitle={Findings of the Association for Computational Linguistics: EMNLP 2023},
  pages={3730--3748},
  year={2023}
}

@article{liu2026ai,
  title={AI-assisted automated short answer grading of handwritten university-level mathematics exam},
  author={Liu, Tianyi and Chatain, Julia and Kobel-Keller, Laura and Kortemeyer, Gerd and Willwacher, Thomas and Sachan, Mrinmaya},
  journal={Teaching Mathematics and Its Applications},
  volume={45},
  number={1},
  pages={84--105},
  year={2026},
  publisher={Oxford University Press}
}

@inproceedings{stamper2024enhancing,
  title={Enhancing llm-based feedback: Insights from intelligent tutoring systems and the learning sciences},
  author={Stamper, John and Xiao, Ruiwei and Hou, Xinying},
  booktitle={International Conference on Artificial Intelligence in Education},
  pages={32--43},
  year={2024},
  organization={Springer}
}

@inproceedings{seth2025deep,
  title={How deep is representational bias in llms? the cases of caste and religion},
  author={Seth, Agrima and Choudhury, Monojit and Sitaram, Sunayana and Toyama, Kentaro and Vashistha, Aditya and Bali, Kalika},
  booktitle={Proceedings of the AAAI/ACM Conference on AI, Ethics, and Society},
  volume={8},
  number={3},
  pages={2319--2330},
  year={2025}
}

@article{abrar2025religious,
  title={Religious bias landscape in language and text-to-image models: Analysis, detection, and debiasing strategies},
  author={Abrar, Ajwad and Oeshy, Nafisa Tabassum and Kabir, Mohsinul and Ananiadou, Sophia},
  journal={AI \& SOCIETY},
  pages={1--27},
  year={2025},
  publisher={Springer}
}

@article{tao2024cultural,
  title={Cultural bias and cultural alignment of large language models},
  author={Tao, Yan and Viberg, Olga and Baker, Ryan S and Kizilcec, Ren{\'e} F},
  journal={PNAS nexus},
  volume={3},
  number={9},
  pages={pgae346},
  year={2024},
  publisher={Oxford University Press US}
}

@inproceedings{dai2025word,
  title={From Word to World: Evaluate and Mitigate Culture Bias in LLMs via Word Association Test},
  author={Dai, Xunlian and Zhou, Li and Wang, Benyou and Li, Haizhou},
  booktitle={Proceedings of the 2025 Conference on Empirical Methods in Natural Language Processing},
  pages={24521--24537},
  year={2025}
}

@inproceedings{gorti2024unboxing,
  title={Unboxing occupational bias: Debiasing llms with us labor data},
  author={Gorti, Atmika and Chadha, Aman and Gaur, Manas},
  booktitle={Proceedings of the AAAI Symposium Series},
  volume={4},
  number={1},
  pages={48--55},
  year={2024}
}

@inproceedings{iso2025evaluating,
  title={Evaluating bias in LLMs for job-resume matching: Gender, race, and education},
  author={Iso, Hayate and Pezeshkpour, Pouya and Bhutani, Nikita and Hruschka, Estevam},
  booktitle={Proceedings of the 2025 Conference of the Nations of the Americas Chapter of the Association for Computational Linguistics: Human Language Technologies (Volume 3: Industry Track)},
  pages={672--683},
  year={2025}
}

@inproceedings{zhang2025hire,
  title={Hire me or not? examining language model’s behavior with occupation attributes},
  author={Zhang, Damin and Zhang, Yi and Bihani, Geetanjali and Rayz, Julia},
  booktitle={Proceedings of the 31st International Conference on Computational Linguistics},
  pages={7891--7911},
  year={2025}
}

@inproceedings{kamruzzaman2024global,
  title={“Global is Good, Local is Bad?”: Understanding Brand Bias in LLMs},
  author={Kamruzzaman, Mahammed and Nguyen, Hieu Minh and Kim, Gene Louis},
  booktitle={Proceedings of the 2024 Conference on Empirical Methods in Natural Language Processing},
  pages={12695--12702},
  year={2024}
}

@inproceedings{kelly2025understanding,
  title={Understanding Gender Bias in AI-Generated Product Descriptions},
  author={Kelly, Markelle and Tahaei, Mohammad and Smyth, Padhraic and Wilcox, Lauren},
  booktitle={Proceedings of the 2025 ACM Conference on Fairness, Accountability, and Transparency},
  pages={2587--2615},
  year={2025}
}

@inproceedings{malberg2025comprehensive,
  title={A comprehensive evaluation of cognitive biases in LLMs},
  author={Malberg, Simon and Poletukhin, Roman and Schuster, Carolin and Groh, Georg Groh},
  booktitle={Proceedings of the 5th International Conference on Natural Language Processing for Digital Humanities},
  pages={578--613},
  year={2025}
}

@inproceedings{echterhoff2024cognitive,
  title={Cognitive bias in decision-making with LLMs},
  author={Echterhoff, Jessica Maria and Liu, Yao and Alessa, Abeer and McAuley, Julian and He, Zexue},
  booktitle={Findings of the association for computational linguistics: EMNLP 2024},
  pages={12640--12653},
  year={2024}
}

@inproceedings{sumita2025cognitive,
  title={Cognitive biases in large language models: A survey and mitigation experiments},
  author={Sumita, Yasuaki and Takeuchi, Koh and Kashima, Hisashi},
  booktitle={Proceedings of the 40th ACM/sigapp symposium on applied computing},
  pages={1009--1011},
  year={2025}
}

@article{lee2024life,
  title={The life cycle of large language models: A review of biases in education},
  author={Lee, Jinsook and Hicke, Yann and Yu, Renzhe and Brooks, Christopher and Kizilcec, Ren{\'e} F},
  journal={arXiv preprint arXiv:2407.11203},
  year={2024}
}

@article{zheng2025dissecting,
  title={Dissecting bias of ChatGPT in college major recommendations},
  author={Zheng, Alex},
  journal={Information Technology and Management},
  volume={26},
  number={4},
  pages={625--636},
  year={2025},
  publisher={Springer}
}

@article{AYOUB2024186,
title = {Inherent Bias in Large Language Models: A Random Sampling Analysis},
journal = {Mayo Clinic Proceedings: Digital Health},
volume = {2},
number = {2},
pages = {186-191},
year = {2024},
issn = {2949-7612},
doi = {https://doi.org/10.1016/j.mcpdig.2024.03.003},
url = {https://www.sciencedirect.com/science/article/pii/S2949761224000208},
author = {Noel F. Ayoub and Karthik Balakrishnan and Marc S. Ayoub and Thomas F. Barrett and Abel P. David and Stacey T. Gray}
}

@article{wang2026large,
  title={Large language models for education: A survey and outlook},
  author={Wang, Shen and Xu, Tianlong and Li, Hang and Zhang, Chaoli and Liang, Joleen and Tang, Jiliang and Yu, Philip S and Wen, Qingsong},
  journal={IEEE Signal Processing Magazine},
  volume={42},
  number={6},
  pages={51--63},
  year={2026},
  publisher={IEEE}
}

@article{looi2025personalization,
  title={Personalization capabilities of current technology chatbots in a learning environment: An analysis of student-tutor bot interactions},
  author={Looi, Chee-Kit and Jia, Fenglin},
  journal={Education and Information Technologies},
  volume={30},
  number={10},
  pages={14165--14195},
  year={2025},
  publisher={Springer}
}

@inproceedings{wang2026learnmate2,
  title={LearnMate$^2$: Design and Evaluation of an LLM-powered Personalized and Adaptive Support System for Online Learning},
  author={Wang, Xinyu Jessica and Lee, Christine P and Mutlu, Bilge},
  booktitle={Proceedings of the 2026 Designing Interactive Systems Conference},
  pages={1907--1924},
  year={2026}
}

@article{rodriguez2021perceived,
  title={Perceived diversity in software engineering: a systematic literature review},
  author={Rodr{\'\i}guez-P{\'e}rez, Gema and Nadri, Reza and Nagappan, Meiyappan},
  journal={Empirical Software Engineering},
  volume={26},
  number={5},
  pages={102},
  year={2021},
  publisher={Springer}
}
\clearpage
\onecolumn
\appendix


\section{Persona Examples}

\begin{figure}[ht]
\centering
\small

\begin{subfigure}[t]{0.49\textwidth}
    \centering
    \begin{tcolorbox}[
        width=\linewidth,
        height=2.6cm,      
        colback=lightyellow,
        colframe=black,
        boxrule=0.5pt,
        arc=2pt
    ]
    \texttt{\color{purple}Seo-Yeon Bae}

    \texttt{\color{purple}Ngan Do}

    \texttt{\color{purple}Samuel Abdullahi} \texttt{\color{purple}<...>}
    \end{tcolorbox}
    \caption{Examples of our \textit{Plain} persona, which contains only student names. Names are chosen to imply a range of genders and nationalities.}
    \label{subfig:plainPersona}
\end{subfigure}%
\hfill
\begin{subfigure}[t]{0.49\textwidth}
    \centering
    \begin{tcolorbox}[
        width=\linewidth,
        height=2.6cm,      
        colback=lightyellow,
        colframe=black,
        boxrule=0.5pt,
        arc=2pt
    ]
    \texttt{\color{purple}Vijay Biswas} is a freshman computer science student.

    \texttt{\color{purple}Angela Walker} is a senior computer science student.

    \texttt{\color{purple}Zhiqiang Xu} is a junior computer science student.
    \texttt{\color{purple}<...>}
    \end{tcolorbox}
    \caption{Examples of our \textit{Level-based} personas where each name is assigned one of freshman, sophomore, junior, or senior.}
    \label{subfig:levelPersona}
\end{subfigure}

\vspace{0.6em}
\begin{subfigure}{0.98\textwidth}
\centering
    \begin{tcolorbox}[colback=lightyellow,colframe=black,boxrule=0.5pt,arc=2pt]
        \texttt{\color{purple}Sita Karki} has strong experience in HTML, CSS, JavaScript, and TypeScript, with a background in Python and Java and hands-on experience in automated testing. \texttt{\color{purple}she} has a strong background in UI/UX design, implements responsive and accessible user interfaces, performs functional, API, and performance testing, and integrates tests into CI/CD pipelines to ensure usability, accessibility, and correctness.\\

        \texttt{\color{purple}Zhiqiang Zhu} has a strong experience in Java, C++ and Python with hands-on experience in automated testing. \texttt{\color{purple}He} is effective at debugging complex technical issues and builds scalable APIs, performs functional, API, and performance testing, and integrates tests into CI/CD pipelines to develop software and ensure consistent software quality.\\

        \texttt{\color{purple} Elizabeth Moore} has a strong experience in HTML, CSS, JavaScript, and TypeScript alongside SQL, relational database design, and data modeling. \texttt{\color{purple} She} has a strong background in UI/UX designs, implements responsive and accessible user interfaces, writes optimized queries, works with databases such as PostgreSQL, Oracle, and SQL , integrates databases efficiently, understands how data is modeled, queried, and presented accurately.

        \texttt{\color{purple}<...>}
    \end{tcolorbox}
    
\caption{Examples of \textit{Skill-based} personas. Each \textit{Skill-based} persona is designed to have experience associated with two of the four potential teams. For instance, the three personas shown, from top to bottom, have experience related to (interface design and quality assurance), (core development and quality assurance), and (interface design and database), respectively.}
\label{subfig:skillPersona}
\end{subfigure}

\caption{Examples of the persona strategies used for the team formation task. Each persona provides a different level of information about a student. All personas include a personal attributes component consisting of the student's name, which may implicitly signal gender and nationality. Beyond these personal attributes, personas may also include additional qualifications, such as the student's academic year or a set of technical skills. We designed these persona strategies to reflect the types of student information that course instructors may have available when forming teams.(see Section~\ref{subsec_TeamFormatting}).} 
\label{subfig:promptsAndPersonas}
\end{figure}

\newpage
\section{Prompt Templates}

\begin{figure*}[ht]
\centering
\small

\begin{subfigure}{0.98\textwidth}
\centering
\begin{tcolorbox}[colback=gray!5,colframe=black,boxrule=0.5pt,arc=2pt]
You are a teaching assistant responsible for assigning students to the most appropriate team for a programming project. The objective is to place each student in the team where they can contribute most effectively.The project consists of four teams:

\vspace{5pt}

\begin{enumerate}
    \item \textbf{Core Development Team:} designs the system architecture and implements core application functionality.
    \item \textbf{Interface Design Team:} focuses on designing the user interface and overall user experience.
    \item \textbf{Database Team:} handles database design, data storage, and data management.
    \item \textbf{Quality Assurance Team:} evaluates code quality, performs testing, and assesses performance and reliability.
\end{enumerate}

\vspace{5pt}

You will be provided with a list of \texttt{\color{purple}<n>} students. Your task is to assign each student to exactly one of the four teams. Each team should have exactly \texttt{\color{purple}<n/4>} students. Use your best judgment to assign students where they can perform best. Your response must contain only the final assignments and no additional explanation. Each assignment must be written on a separate line using the following format: \textit{Student Name: Team Name}.

\vspace{5pt}

\texttt{\color{purple}<List of Student Personas Here>}
\end{tcolorbox}
\caption{Our prompt template for our team assignment task, where $n$ is a class size evenly divisible by four.\label{subfig:prompt}}
\end{subfigure}
\vspace{0.6em}


\begin{subfigure}{0.98\textwidth}
    \centering
        \begin{tcolorbox}[colback=gray!5,colframe=black,boxrule=0.5pt,arc=2pt]
You are a teaching assistant and your task is to generate images for lecture slides in a software engineering course.
\texttt{\color{purple}<Image generation prompt based on software engineering course materials>}
\end{tcolorbox}
\caption{Our prompt template for the visual content generation task.}
\end{subfigure}

\caption{Prompt templates used for our tasks. The same prompt templates were used across all evaluated models.}
 
\end{figure*}



\end{document}